\documentclass[twocolumn]{revtex4}
\newcommand{\be}{\begin{eqnarray}}
\newcommand{\ee}{\end{eqnarray}}
\newcommand{\ket}[1]{\ensuremath{\left| {#1} \right>}}
\newcommand{\bra}[1]{\ensuremath{\left< {#1} \right|}}

\newcommand{\create}{\ensuremath{{\,\hat{a}^{\dagger}}}}
\newcommand{\destroy}{\ensuremath{\,\hat{a}}}

\usepackage{latexsym}
\usepackage{graphics}
\usepackage{amssymb}
\usepackage{multirow}
\usepackage{textcomp}

\begin{document}	
\title{Observation of quantum interference between separated mechanical oscillator wavepackets}

\author{ D.~Kienzler, C. Fl{\"u}hmann, V. Negnevitsky, H.-Y. Lo,  M. Marinelli, D. Nadlinger, J.~P.~Home }\email{jhome@phys.ethz.ch}

\address{Institute for Quantum Electronics, ETH Z\"urich, Otto-Stern-Weg 1, 8093 Z\"urich, Switzerland}

%\ead{jhome@phys.ethz.ch}

\begin{abstract}
%\linenumbers
The ability of matter to be superposed at two different locations while being intrinsically connected by a quantum phase is among the most counterintuitive predictions of quantum physics. While such superpositions have been created for a variety of systems \cite{13Wineland,91Kasevich}, the in-situ observation of the phase coherence has remained out of reach. Using a heralding measurement on a spin-oscillator entangled state,
we project a mechanical trapped-ion oscillator into a superposition of two spatially separated states, a situation analogous to Schr{\"o}dinger's cat \cite{BkHaroche}. Quantum interference is clearly observed by extracting the occupations of the energy levels. For larger states, we encounter problems in measuring the energy distribution, which we overcome by performing the analogous measurement in a squeezed Fock basis with each basis element stretched along the separation axis. Using 8~dB of squeezing we observe quantum interference for cat states with phase space separations of $\Delta \alpha = 15.6$, corresponding to wavepackets with a root-mean-square extent of $7.8$~nm separated by over 240~nm. We also introduce a method for reconstructing the Wigner phase-space quasi-probability distribution using both squeezed and non-squeezed Fock bases. We apply this to a range of negative parity cats, observing the expected interference fringes and negative values at the center of phase space. Alongside the fundamental nature of these large state superpositions, our reconstruction methods facilitate access to the large Hilbert spaces required to work with mesoscopic quantum superpositions, and may be realized in a wide range of experimental platforms \cite{15Wollman, 11Teufel, 15Pirkkalainen}.
\end{abstract}

\pacs{pacs}
\maketitle

%\linenumbers

The rules of quantum mechanics give rise to the prediction that systems can exist in a superposition of two macroscopically distinct quantum states, connected by a fixed relationship known as the quantum phase. This is illustrated by the Schr{\"o}dinger's ``cat'' thought experiment, in which a cat is envisioned as being simultaneously dead and alive, a situation which has no counterpart in our classically familiar world. The key distinction which separates the quantum superposition from a classical mixture is the phase relationship between the two distinct parts of the superposition. An approximation to such a situation is provided by superposed ``classical'' coherent states of oscillators which are macroscopically distinct at large amplitudes. Such ``cat'' states have been realized in the oscillations of trapped atomic ions \cite{96Monroe, 07McDonnell, 05Haljan, 13Wineland, 15Lo}, and for the electromagnetic field \cite{13Haroche, 13Vlastakis}. While for the latter the direct phase relationship between the two states has been observed, for massive particles (such as trapped ions and matter-wave interferometers \cite{91Kasevich}) in which spatial superpositions have been created no in-situ measurement has been performed. Quantum coherence has instead been verified by bringing the two separated wavepackets together so that they spatially overlap, and observing the resulting revival of coherence \cite{96Monroe, 15Lo}. For large cat sizes, both the mean and the uncertainty in energy of the superposed wavepackets is increased. As viewed from the energy eigenbasis, the states occupy an increasingly large Hilbert space, and thus become progressively harder to characterize. This provides an additional challenge to experiments which seek to probe cat states in the mesoscopic regime.

In this Letter, we use an in-sequence spin measurement on a spin-motion entangled state to project out and  herald a superposition of two coherent mechanical oscillator states of opposite phase. We perform measurements which directly observe the interference of the two spatially separated wavepackets through the effect on the occupation of the energy eigenstates. For $\alpha > 5$ the standard analysis method, which is based on an energy eigenstate decomposition, has poor signal to noise. We overcome this limitation by performing an analogous eigenstate decomposition in a squeezed Fock basis in which the mean quantum number of the cat is reduced substantially. Using a Fock basis with 8~dB of squeezing, we are able to observe quantum interference for phase-space separations of $\Delta \alpha = 15.6$, which correspond to a direct measurement of interference between wavepackets separated by 240~nm with a root-mean-square extent of $7.8$~nm.  Adding an extra term to the probe Hamiltonian, we displace the analysis basis, providing a method for reconstructing the Wigner function of the oscillator state. We demonstrate a full reconstruction using the non-squeezed basis for a cat with $\alpha = 2.1$, and take additional slices through the phase space for cats with $\alpha = 4.25$ and $5.9$, making use of a squeezed analysis basis with 7~dB of squeezing for the larger state.

\begin{figure}[t]
	\resizebox{0.48\textwidth}{!}{\includegraphics{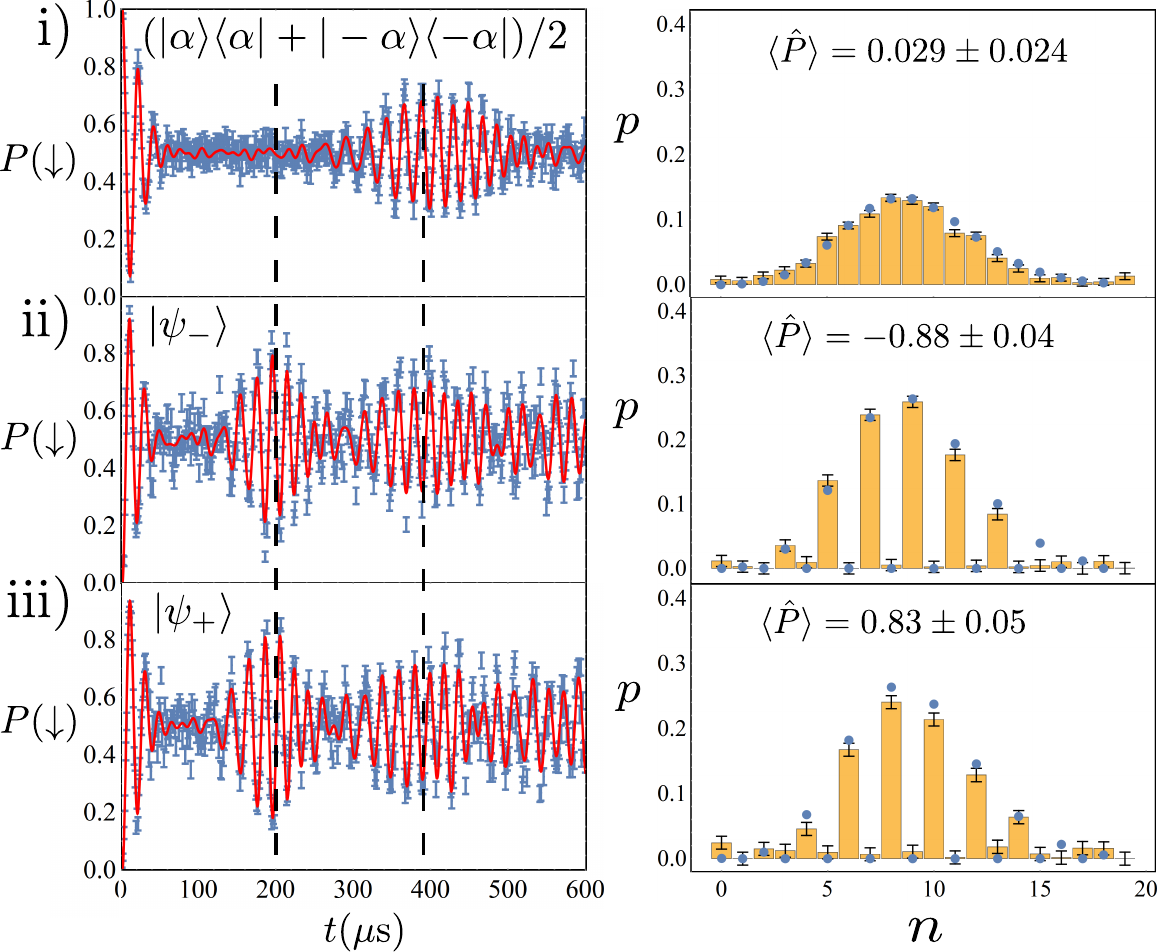}}
	\caption{Experimental data from measurements of $P(\downarrow, t_p)$ using $\hat{H}_r$ (left). These are fitted using equation \ref{eq:spinpops} to obtain the motional energy eigenstate populations shown on the right. The data and populations include cases where the creation of state $\ket{\psi_{\rm ent}}$ is followed by an analysis which is performed after i) repumping the spin to $\ket{\downarrow}$, ii) detection of the ion in the state $\ket{\uparrow}$, producing the state $\ket{\psi_-}$ and iii) detection of the ion in the state $\ket{\uparrow}$ after applying a spin flip using the carrier transition, which produces $\ket{\psi_+}$. The vertical dashed lines indicate the revival times $t_r$ and $t_{r, \rm mix}$. The reconstructed population distributions show clearly the effect of the post-selected measurement for the latter two cases. The blue points correspond to the ideal cases for $\alpha = 3$. Spin populations are the result of 250 repeats of the full experimental sequence, which corresponds to roughly 125 analysis detections for the post-selected cases.
	Error bars are estimated from quantum projection noise. The population errors are given as s.e.m..}
	\label{fig:projection}
	%Data used and output plots are in in Y2015_ProjectedCats\Data Analysis\20151107 Cat analysis, along with analysis_PaperData.nb which produced the basic plots.
\end{figure}

The oscillator cat states are experimentally generated starting from an ion prepared in the internal state $\ket{\downarrow}$ and the oscillator ground state (more details of experimental methods can be found in \cite{Methods}). We then apply an internal-state dependent force using a Hamiltonian $\hbar \Omega\hat{\sigma}_x (\create + \destroy)/2$, where $\hat{\sigma}_x = \ket{+}\bra{+} - \ket{-}\bra{-}$ with $\ket{\pm} \equiv (\ket{\downarrow} \pm \ket{\uparrow})/\sqrt{2}$, and $\Omega$ is a constant. In our experiments this is well approximated by simultaneously driving the red and blue motional sidebands of the internal state transition $\ket{\downarrow}\leftrightarrow\ket{\uparrow}$  \cite{96Monroe} (corrections due to higher order terms in the laser-ion interaction are relatively small - see \cite{Methods} for more information). After applying this Hamiltonian for a duration $t$, ideally the state of the ion can be written as a spin-motion entangled state
\be
\ket{\psi_{\rm ent}} = \frac{1}{\sqrt{2}}\left(\ket{+}\ket{\alpha} + \ket{-}\ket{-\alpha}\right)
\ee
where $\alpha =  -i \Omega t/2$ relates the amplitude of oscillation $z$ to the  r.m.s. extent of the ground state wavefunction $z_0$ via $z = 2 \alpha z_0$. This is the standard Schr{\"o}dinger's cat state which has been produced in earlier work with trapped ions \cite{96Monroe, 07McDonnell, 05Haljan, 13Wineland}, in which the superposed motional wavepackets are separated in phase space by $\Delta\alpha = 2 \alpha$ but are entangled with the ion's internal state. $\ket{\psi_{\rm ent}}$ can be written in the $\ket{\uparrow}$, $\ket{\downarrow}$ basis as
\be
\ket{\psi_{\rm ent}} = \frac{1}{\sqrt{2}}\left(\ket{\uparrow} \ket{\psi_-}  + \ket{\downarrow}\ket{\psi_+}\right)
\ee
where $\ket{\psi_{\pm}} \equiv (\ket{\alpha }\pm\ket{-\alpha})/\sqrt{2}$. A measurement of the internal state therefore projects the motional state into a superposition of two out-of-phase oscillations, with the quantum phase relating the two depending on the measurement result. Detection of the internal state involves applying a resonant laser which scatters many photons for the $\ket{\downarrow}$ state, but none for the $\ket{\uparrow}$ state \cite{98Wineland2}. The former results in photon recoil which destroys the motional state, thus we make an in-sequence decision to only proceed with the analysis of the motional state when the internal state is found to be  $\ket{\uparrow}$. The state $\ket{\psi_{+}}$ ($\ket{\psi_-}$) contains only even (odd) energy eigenfunctions, due to the quantum interference between the two coherent states $\ket{\pm \alpha}$. The state populations $p(\ket{n})$ are extracted by observing the spin projection as a function of the duration $t_p$ of a probe Hamiltonian $\hat{H}_r = (\hbar \Omega_r/2) \left[\create \hat{\sigma}_- + {\rm h.c.}\right]$. Experimentally this Hamiltonian is realized by driving the red sideband transition $\ket{\uparrow, n}\leftrightarrow\ket{\downarrow, n + 1}$. The probability of observing spin $\ket{\downarrow}$ after the pulse follows
\be
P(\downarrow, t_p) = \frac{1}{2}\sum_n p(\ket{\phi_n}) \left( 1 - \gamma(t) \cos(\Omega_{n, n + 1} t_p/2) \right) \ , \label{eq:spinpops}
\ee
where  $p(\ket{\phi_n})$ is the probability that the motion started in the Fock state $\ket{\phi_n}$ prior to the probe pulse. When using the red-sideband probe these Fock states correspond to the energy eigenstates $\ket{\phi_n} = \ket{n}$ and the Rabi frequencies $\Omega_{n, n + 1} = \Omega_r M_{n}$ scale with $n$ according to the motional matrix elements $M_n = \bra{n+1}e^{i \eta (\create + \destroy)}\ket{n}$. For small values of the dimensionless
Lamb-Dicke parameter $\eta$ these scale as $\sqrt{n +1}$ \cite{03Leibfried2}. $\gamma(t)$ accounts for decay in the coherence of the whole system during the probe pulse. In fitting data we use a phenomenological exponential decay $\gamma(t) = e^{-\Gamma t}$ (more details regarding the fits can be found in \cite{Methods}). Data obtained using this method is shown for three states with $|\alpha| \simeq 3$ in figure \ref{fig:projection}, alongside the motional populations obtained from fits of equation \ref{eq:spinpops} to the data. In the first, we do not perform a post-selected measurement, but rather probe the motional state populations after repumping to $\ket{\downarrow}$, resulting in the density matrix $\hat{\rho}_m = (\ket{\alpha}\bra{\alpha} + \ket{-\alpha}\bra{-\alpha})/2$ which has a Poisson population distribution of energy eigenstates (this leaves the ion in $\ket{\downarrow}$, and thus in contrast to all other measurements in this paper we probe using the blue sideband $\ket{\downarrow, n}\leftrightarrow\ket{\uparrow, n + 1}$, resulting in a spin probability $P'(\downarrow,t_p) = 1 - P(\downarrow,t_p)$).
The time evolution of the spin populations shows the well-known collapse and revival, with the latter occuring when Rabi oscillations for neighboring values of $n$ come into phase \cite{BkHaroche}. This occurs around $t_{r, {\rm mix}} \simeq 4 \pi/(\Omega_r (\sqrt{\langle n \rangle + 1} - \sqrt{\langle n \rangle})$, corresponding to $t_{r, {\rm mix}} = 386$~$\mu$s for the settings $\Omega_r/(2 \pi) = 31$~kHz and $\bar{n} = |\alpha|^2 = 8.76$ used in our experiment. The results for the mixed state should be compared with cases in which the states are analyzed conditional on the results of a spin measurement of the internal state. 
The first corresponds to analyzing the motional state only when the spin is measured to be $\ket{\uparrow}$, which ideally produces $\ket{\psi_-}$. For the second, we perform  a coherent spin inversion prior to the conditional measurement, which ideally projects the motion into $\ket{\psi_+}$. For both $\ket{\psi_+}$ and $\ket{\psi_-}$ the revival at $t_p = t_{r, {\rm mix}}$ is accompanied by an additional revival in the spin population oscillations at around 198~$\mu$s which results from re-phasing of the contributions from Fock states differing in $n$ by 2. This corresponds to the condition $t_r = 4\pi/(\Omega_r (\sqrt{\langle n \rangle + 2} - \sqrt{\langle n \rangle})$ which for $\langle n \rangle\gg 2$ gives $t_r \simeq 4\pi \sqrt{\langle n \rangle}/\Omega_r \simeq t_{r, {\rm mix}}/2$. We see in the  populations extracted from the fit that the odd and even cat states contain predominantly odd or even number states. We use these populations to extract the parity $\langle \hat{P} \rangle \equiv \sum_n (-1)^n p(\ket{n})$ of the number state distributions, obtaining $0.029\pm0.024$ for the mixture, $-0.88\pm0.04$ for $\ket{\psi_-}$ and $0.83\pm0.05$ for $\ket{\psi_+}$.

\begin{figure}[h!]
\resizebox{0.46\textwidth}{!}{\includegraphics{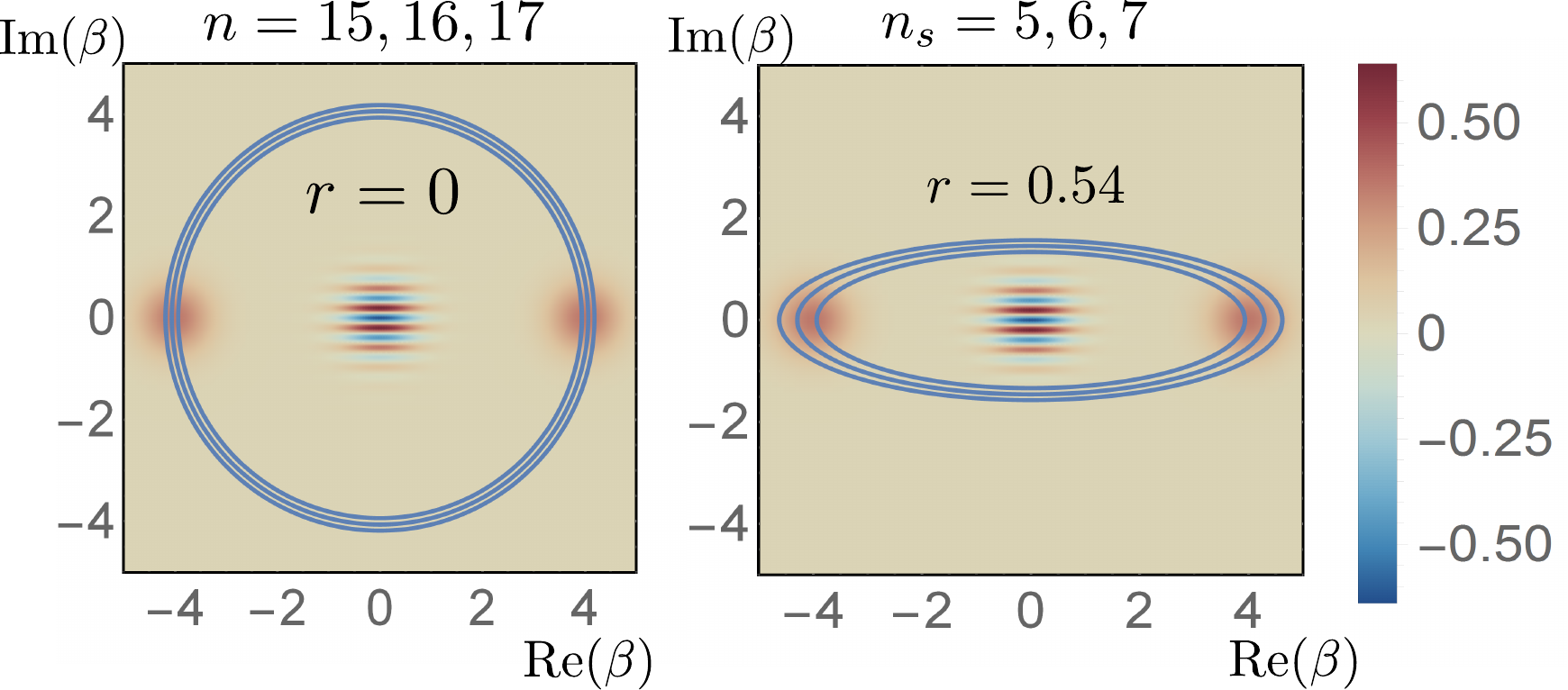}}
\caption{Illustration of a comparison between the energy eigenbasis and the squeezed basis for $\alpha = 4$ and $r = 0.54$. The Wigner function of the cat state is overlayed by lines which indicate the maximal quasi-probability of the three closest Fock states to the mean value of the cat state in the relevant basis. This is approximately given by $\beta  = (n + 1/2)^{1/2}(\cos(\theta) e^r + i \sin(\theta) e^{-r})$ \cite{BkSchleich}. The use of the squeezed basis reduces both the mean value and the variance, simplifying extraction of the motional populations from the spin oscillation data.}
\label{fig:illustrationsqueezed}
\end{figure}

In the methods used above, the visibility of the revival of oscillations in the spin population at time $t_r$ is the key element for diagnosing the parity of the cat. Since the mean Rabi rate scales as $\sqrt{\langle n \rangle + 1}$, the revival time corresponds to approximately $\langle n \rangle$ Rabi oscillations on the motional sideband. In order to observe a significant revival amplitude, this time must be short compared to relevant decoherence times for the cat state, and the stability of the Rabi oscillations must be high enough that $\langle n \rangle$ oscillations are visible for all relevant Fock states. The coherence time of trapped-ion oscillator cats due to motional heating and motional dephasing scales roughly as $1/|\Delta\alpha|^2$ \cite{00Turchette2} and we observe that our spin-motion Rabi oscillations decay with a time-constant which is proportional to the Rabi frequency (see discussion in \cite{Methods}). The combination of these two effects means that we are unable to observe a useful interference feature for a cat with $\Delta\alpha   \gtrsim 10$. To overcome this problem, we introduce an analysis method based on a squeezed Fock state basis $\ket{\phi_{n_s}} = \ket{n_s} \equiv \hat{S}(\xi)\ket{n}$ where the squeezing operator is defined as $\hat{S}(\xi) \equiv e^{(\xi^* \destroy^2 - \xi \create^2)/2}$ with $\xi \equiv r e^{i\phi_s}$ \cite{15Kienzler}. The basic idea is illustrated in figure \ref{fig:illustrationsqueezed}. We choose the squeezing axis perpendicular to the axis separating the two wavepackets of the cat, which corresponds to selecting the phase $\phi_s = 2 \arg{\alpha} + \pi$. The effect of the anti-squeezing results in a mean occupation in the squeezed basis of $\langle n_s\rangle = |\alpha|^2 e^{- 2r} + \sinh^2(r)$, exponentially suppressing the contribution from the displacement at the cost of an additional contribution which can be kept small with an appropriate choice of $r$.  The minimum value of $\langle n_s \rangle$ for a given $\alpha$ is reached for $r_{\rm min} = \ln(4 |\alpha|^2 + 1)/4$. Since the squeezing operator preserves parity, the even (odd) cats consist of only even (odd) number state populations in both bases. The squeezed reconstruction also has the advantage that it is directly sensitive to the phase of the cat, which the standard analysis is not.

\begin{figure}[h!]
\vspace{12pt}
\resizebox{0.48\textwidth}{!}{\includegraphics{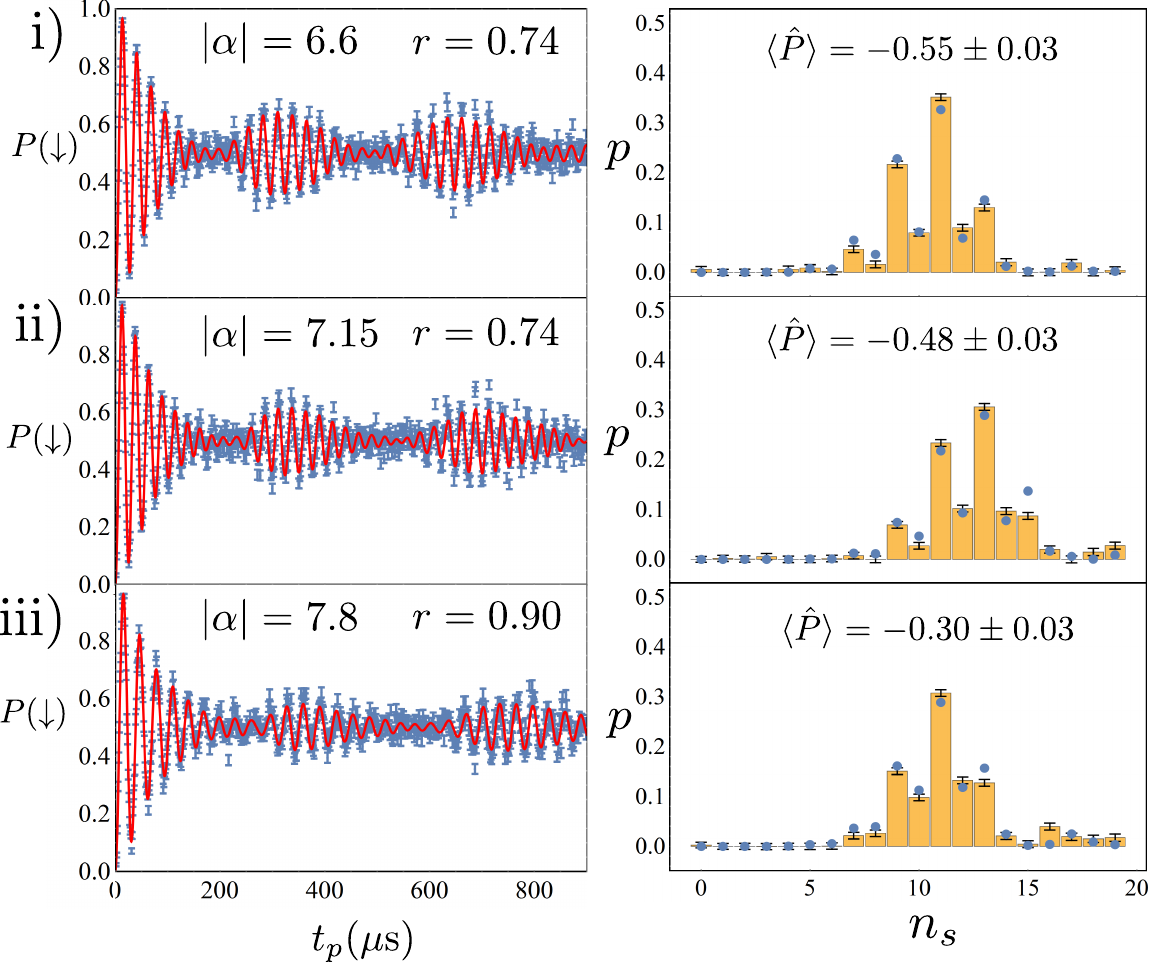}}
\caption{Experimental data for the spin evolution $P(\downarrow, t_p)$ obtained while probing in the squeezed basis with $\hat{H}_s$ (shown on left). These are fitted using equation \ref{eq:spinpops} to obtain the squeezed Fock state populations (shown on right). The parameters for the cat size and the squeezing amplitude are given, along with the parity value obtained from the extracted populations. The blue points in the Fock state population plots are the populations obtained from a fit using a weighted mixture of $\ket{\psi_+}$ and $\ket{\psi_-}$. Experimental data points are an average of 750 repeats of the full experimental sequence (1000 for iii)), which corresponds to roughly 375 (500) analysis detections for the post-selected cases. Error bars are estimated from quantum projection noise. Errors on population and parity estimates are given as the s.e.m..}
\label{fig:squeezed}
%Data used and output plots are in in Y2015_ProjectedCats\Data Analysis\20151107 Cat analysis, along with analysis_PaperData.nb which produced the basic plots.
\end{figure}

To measure our states in the squeezed basis, we substitute the Hamiltonian $\hat{H}_s \equiv (\hbar \Omega_s/2)\left[(\destroy + \tanh(r)  e^{i\phi_s}\create)\hat{\sigma}_- + {\rm h.c.}\right]$ for $\hat{H}_r$ (for calibration of the phases, see \cite{Methods}). The time evolution of the spin populations follows equation \ref{eq:spinpops} with the relevant number states becoming those of the squeezed basis. In the Lamb-Dicke regime the scaling of the Rabi frequencies is again $\sqrt{n_s + 1}$, though in fitting the data we use values for the matrix elements which include higher order terms. Data for cat states with $|\alpha| = 6.6, 7.15$ and $7.8$ are shown in figure \ref{fig:squeezed}. The optimal choice of $r$ for minimizing $\langle n_s \rangle$ was not used, because we observe that for larger $r$ the Rabi oscillations of the squeezed Hamiltonian dephase, which impedes the signal. In addition, we do not gain the full speed-up in the revival time because in our setup $\Omega_{s} < \Omega_r$. Nevertheless, for our largest cat the number of Rabi oscillations at which the revival occurs is reduced from 60 to 11, which is essential for observing the interference. By fitting equation \ref{eq:spinpops} with freely floated number state populations, we obtain the results shown in figure \ref{fig:squeezed}, from which we extract parities of $\langle \hat{P} \rangle = -0.55\pm0.03$, $-0.48\pm0.03$ and $-0.30\pm0.03$ for $|\alpha| = 6.6$, $|\alpha| = 7.15$ and $|\alpha| = 7.8$ respectively. Also shown in figure \ref{fig:squeezed} are the populations obtained from a fit to the experimental $P(\downarrow, t_p)$ using a model of the motional populations which is derived from a weighted sum of the even and odd cats $\hat{\rho}_{\rm mix} = \xi_{\rm mix}\ket{\psi_-}\bra{\psi_-} + (1 - \xi_{\rm mix})\ket{\psi_+}\bra{\psi_+}$. The close match between the theory and experiment indicate that the primary decoherence mechanism mixes the two cat states, which is compatible with heating of the ion due to fluctuations in the electric field at frequencies close to the ion's oscillation frequency \cite{00Turchette, 00Turchette2}.

The Wigner function is a phase space quasi-probability distribution which plays an important role in visualizing and characterizing oscillator states \cite{BkSchleich}. It can be related to the expectation value of the parity operator for the populations of displaced number states with displacement $\beta$ as $W(\beta) = 2/\pi \langle \hat{P}(\beta) \rangle$  with $\langle \hat{P}(\beta) \rangle = \sum_n (-1)^n p(\hat{D}(\beta)\ket{n})$ and $\hat{D}(\beta) \equiv e^{\beta \create - \beta^* \destroy}$ the displacement operator \cite{96Leibfried, 13Vlastakis}. This relationship has previously been used to experimentally reconstruct Fock, coherent and thermal states of a trapped ion oscillator using a method that involves displacing the state by $-\beta$ followed by extraction of the populations in the energy eigenstate basis \cite{96Leibfried}. Rather than taking this approach, we obtain the populations of the oscillator states directly in the displaced basis, by adding a displacement term $\hat{H}_d(\beta) = \hbar \Omega/2 (\beta^* \hat{\sigma}_+ + \beta \hat{\sigma}_-)$ to the Hamiltonian used to probe the state \cite{15Kienzler}. We do this for both $\hat{H}_r$ and $\hat{H}_s$ by adding a laser component resonant with the carrier transition $\ket{\uparrow}\leftrightarrow\ket{\downarrow}$, with $\Omega$ chosen to equal $\Omega_r$ and $\Omega_s$ for each case. Under the action of the modified Hamiltonian, the spin population dynamics  follow equation \ref{eq:spinpops}, but with the relevant probabilities being those of the displaced number states $\ket{\phi_n} = \hat{D}(\beta) \ket{n}$ or the displaced squeezed number states $\hat{S}(\xi)\hat{D}(\beta) \ket{n}$ \cite{15Kienzler, Methods}. In the Lamb-Dicke regime the matrix elements scale as $\sqrt{n+1}$, though we take account of deviations from this form in fitting experimental results.

\begin{figure}[t]
\vspace{12pt}
\resizebox{0.48\textwidth}{!}{\includegraphics{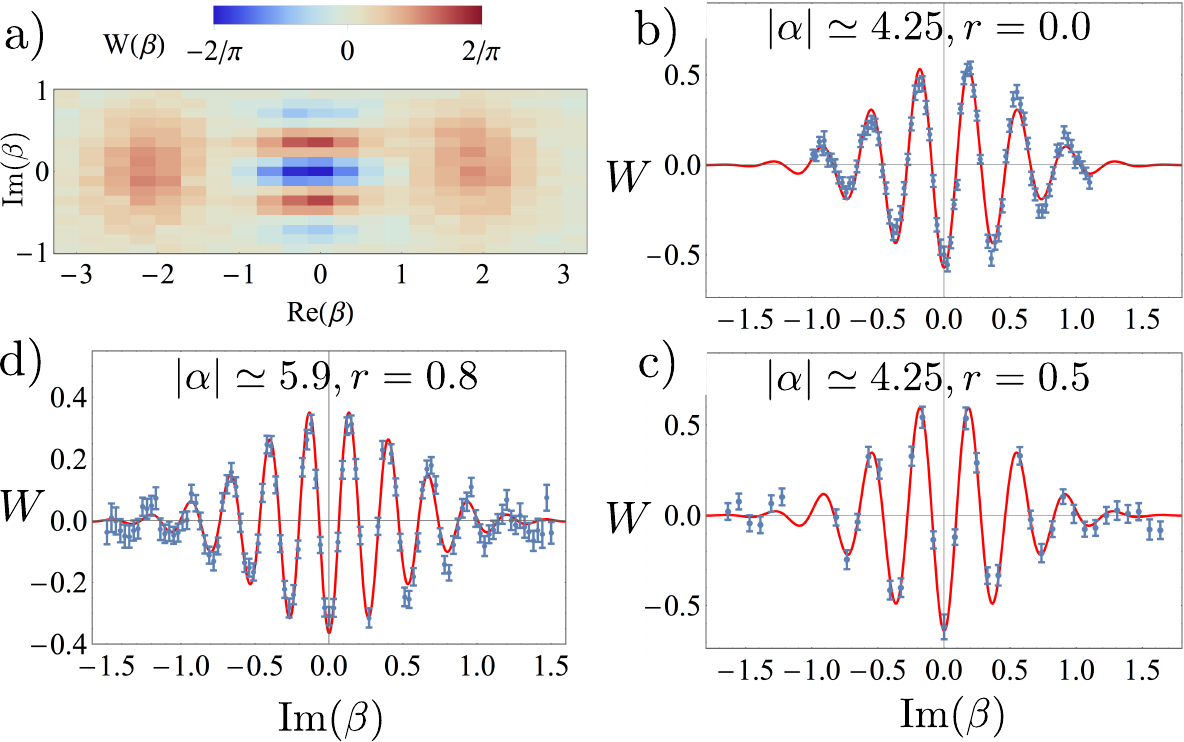}}
\caption{a) Reconstructed Wigner function for a cat with $\alpha \simeq 2.0$, extracted using 289 settings of the displaced Fock state basis. b) Extracted values of the Wigner function on the imaginary axis in phase space for a cat with $\alpha = 4.2$, again extracted using a displaced  basis. c) Wigner function reconstruction for $\alpha \simeq 4.3$ performed using a displaced and squeezed basis with $r = 0.5$. d) Similar extraction for a cat with $\alpha \simeq 6$ performed using a displaced and squeezed basis with $r = 0.8$. In b-d), the fits are a theoretical form which can be derived from the Wigner function of a well-separated cat state (for further details, see \cite{Methods}). Errors bars are given as the s.e.m.. The ${\rm Im}(\beta)$ axis is extracted from periodic calibration scans, and we observe drifts in the scaling at the 10\% level over a day. Thus there is some uncertainty in the values of ${\rm Im}(\beta)$ due to drifts between calibrations.}
\label{fig:wigner}
%Data used and output plots are in in Y2015_ProjectedCats\Data Analysis\20151107 Cat analysis, along with analysis_PaperData.nb which produced the basic plots.
\end{figure}

The reconstructed Wigner function for an odd cat state with $\alpha \simeq 2.1$ is shown in figure \ref{fig:wigner}, based on number state population extraction using equation \ref{eq:spinpops} on a grid of $17\times21$ values of $\beta$. This shows clearly the expected features, including the two separated peaks corresponding to the coherent state wavepackets as well as the interference fringes close to $\beta = 0$. Results are also shown for cuts along the imaginary axis of phase space for a cat with $\alpha \simeq 4.25$ extracted using measurements in both the energy eigenstate basis and in a squeezed basis with $r = 0.5$. We fit the functional form $f(x) = 2/\pi A e^{-2 x^2} \cos(4 \alpha x)$ with $x = {\rm Im}(\beta)$ \cite{BkHaroche}, and extract $\alpha = 4.21\pm0.02$ and $A = 0.90\pm0.02$ for the unsqueezed basis and $\alpha = 4.25\pm0.02$ and $A = 1.00\pm0.03$ for the squeezed basis. The data acquisition time per point is considerably longer for the non-squeezed basis, because extracting the multiple closely spaced high frequency components in the spin oscillations requires a higher sampling frequency and longer probe times. Also shown is similar data for a cat with $\alpha \simeq 5.9$ which was taken using a displaced-squeezed probe Hamiltonian with $r = 0.8$. In this case the fitted curve gives $\alpha = 5.86\pm0.02$ and $A = 0.57\pm0.01$. In all cases the interference fringes which result from the phase relationship of the quantum superposition are clearly visible.

The ability to project the ion into a superposition of different locations and to directly measure the interference of these states allows probing of the cat coherence free from entanglement with the microscopic spin, which is advantageous for probing the limits of quantum coherence in these systems.
In the largest cat observed above, the wavepackets are separated by more than the diffraction limit of $\lambda/2 = 199$~nm of the fluorescence light from the ion. This could allow the two different locations to be resolved with an optical microscope. The use of a squeezed basis in tomographic reconstructions may provide advantages for reconstructing a range of quantum states with large phase space amplitudes. In the work above, it allows the reduction of a state with a mean occupancy of around 60 quanta to one which has a mean of 11. While the main advantage could be expected for states which are extended in one phase space axis, we also anticipate their use for states which contain populations along a larger number of specific directions, with complementary sets of measurements performed along each of these. These methods should be applicable in a range of systems in which similar spin-oscillator couplings are available, including other mechanical oscillators which might be used to probe gravitational or non-linear collapse theories in quantum physics \cite{15Pikovski, 03Bassi}.

%A spatially sensitive detection event synchronized with the oscillation of the ion therefore presents the tantalizing opportunity to resolve the two wavepackets.

%\begin{figure}[hb!]
%\centering
%\resizebox{\textwidth}{!}{
%  }
%\caption{Result of a projective measurement in the $\hat{\sigma}_z$ basis after applying a state-dependent force to an ion prepared in $\ket{\downarrow}$ and in either a squeezed initial state (red, blue) or the ground state of motion (green). For the squeezed state the force was applied along axes corresponding to the squeezed and anti-squeezed quadratures of the state. The data are fitted with $a_1 + a_2 e^{(t/\tau)^2}$, resulting in fitted values $\tau = $ for the ground state, $\tau = $ for the squeezed quadrature and $\tau = $ for the anti-squeezed quadrature. }
%\label{fig:stateoverlap}
%\end{figure}

%It is notable that the long axis data does not reach $P(\uparrow) = 0.5$ at long times (the fitted baseline is 0.55() for this data). We think that this is due to a residual imbalance of Rabi frequencies on the blue and red sideband drives which are used to generate the state-dependent force. Numerical simulations indicate that an imbalance of 0.5\% is sufficient to cause the effect seen in the data. This provides a sensitive tool for the characterization of the light fields, which might be invaluable for multi-qubit gates based on bichromatic light fields \cite{08RoosGate, 99Sorensen, 05Haljan}.

While writing this paper we became aware of parallel work in which a direct parity measurement is used to measure cat states produced in a similar manner using a single motional mode of a two-ion crystal \cite{15Ding}.

\textbf{Acknowledgements:} We thank Lukas Gerster, Ludwig de Clercq and Ben Keitch for contributions to the experimental apparatus. We acknowledge support from the Swiss National Science Foundation under grant no. 200020\_153430, ETH Research Grant under grant no. ETH-18 12-2, and the National Centre of Competence in Research for Quantum Science and Technology (QSIT).\\

\textbf{Author Contributions:} Experimental data was taken by D.K. and C.F., using apparatus and techniques developed by D.K., H.Y.L., C.F., M.M., D.N. and V.N.. Data analysis was performed by D.K., C.F. and J.P.H.. The paper was written by J.P.H. with input from all authors. The work was conceived by J.P.H. and D.K..\\

The authors have no competing financial interests.

\bibliography{./myrefs}

\section*{Methods}

The experiments  make use of a single trapped calcium ion in a micro-fabricated ion trap. The first step of each experimental run involves laser cooling all modes of motion of the ion close to a mean vibrational quantum number of 1 using a combination of Doppler and  Electromagnetically-Induced-Transparency cooling using light at 397, 866~nm \cite{00Roos}. The oscillator for the state engineering is chosen to be the axial mode of motion, in which the ion oscillates with a frequency close to $\omega_z/(2 \pi)$ = 2.08~MHz, which is well resolved from all other modes. This oscillator is cooled further using sideband cooling to ensure a high ground state occupancy. All coherent manipulations make use of the narrow-linewidth transition at 729~nm, isolating a two-state pseudo-spin system which we identify as $\ket{\downarrow} \equiv  \ket{L = 0, J = 1/2, M_J = +1/2}$ and $\ket{\uparrow} \equiv  \ket{L = 2, J = 5/2, M_J = 3/2}$. This transition is resolved by 200~MHz from all other internal state transitions in the applied magnetic field of 11.9~mT. The beam enters the trap at 45 degrees to the $z$ axis of the trap, resulting in a Lamb-Dicke parameter of $\eta \simeq 0.047$ for the axial mode. Optical pumping to $\ket{\downarrow}$ is implemented using a combination of linearly polarized light fields at 854~nm, 397~nm and 866~nm. We read out the internal state of the ion by state-dependent fluorescence observed while applying the laser fields at 397~nm and 866~nm. Motional heating rates from the ground state for an ion in this trap have been measured to be 10~$\pm$~1~quanta\ s$^{-1}$, and the coherence time for the Fock state superposition $(\ket{0}+ \ket{1})/\sqrt{2}$ has been measured to be 32 $\pm$ 3~ms. Spin coherence times have been measured using a Ramsey separated pulse experiment to be 930~$\pm$~20~$\mu$s.

\subsection*{Heralding Detection}
The detection used to decide whether the analysis should be performed or not involves application of lasers at 397~nm and 866~nm, which results in photon scattering from the ion if it is in the ground state $\ket{\downarrow}$, and no scattering for the state $\ket{\uparrow}$. We detect a small fraction  of the photons scattered from the ion, and make a decision based on a pre-set threshold for the detected photon number to decide whether the ion is in $\ket{\uparrow}$ or $\ket{\downarrow}$. We try to keep the detection time short so as to avoid decoherence of the cat. Thus a typical detection time is $75$~$\mu$s. This means that the detection error is around 0.8\% for an ion which is measured to be in $\ket{\downarrow}$, and 0.002\% for an ion which is measured to be in $\ket{\uparrow}$. We think that this error is of similar size to other sources of error in our production of the cat state.

\subsection*{Lamb-Dicke approximation}
The Hamiltonians described in the paper are all given in the Lamb-Dicke approximation which is appropriate in the limit that $\eta^2 (2 n + 1) \ll 1$, where $\eta = k z_0 \cos(\theta)$,
with $\theta$ the angle between the oscillation direction and the laser beam and $k$ the wavevector of the light. The largest states that we produce in  the work in  the paper correspond to $\alpha = 7.8$, or $\langle n \rangle \simeq 60$, and thus we find that the approximation is reasonable, but does not exactly describe our experiments. For this reason we use the full expression for the matrix elements $M_{n, n + 1} = \bra{n + 1} e^{i \eta (\create + \destroy)}\ket{n}$ in fits to sideband data throughout the data analysis. The overall scaling of these is absorbed into the fit Rabi frequency. For the displaced basis we analytically derived expressions which confirm that the same $M_{n, n + 1}$ are appropriate, and thus we use the same scalings as for the energy eigenstate basis in the displaced basis. For the squeezed basis, we haven't been able to derive analytical expressions for the Hamiltonian we apply in the lab, but we have confirmed numerically that the scaling of the Rabi frequencies with $n_s$ are very close to the values estimated from $M_{n_s,n_s+1}$, and the latter have been used in most of the analysis. For the very largest states we use the numerically generated matrix elements for $r = 0.9$ to fit the data, though the correction to the values from $M_{n, n+1}$ is small.

\subsection*{Decay model and observed decoherence.}
In fitting the data using equation \ref{eq:spinpops}, we are forced to make some assumption about the form of decay. This is important because the fitted values of both the decay parameter and the populations which contribute to the parity, are directly related to the  amplitude of the  revival in the spin oscillations at time $t_r$. This correlation means that the parity can be overestimated for data where only the revival at $t_p = t_r$ is included. Therefore it is critical to probe for times which include the revival at $t_{r, {\rm mix}}$, which acts to constrain $\gamma(t)$. In all fits used in this paper we have used $\gamma(t) = e^{-\Gamma t_p}$. An exponential form of decay is often the result of decoherence mechanisms with a short correlation time, which would be appropriate for motional heating \cite{00Turchette}. We use this because it produces the most consistent results between set experimental values and values obtained from fit models throughout our data.

We have also fitted many datasets using Gaussian decay forms $\gamma(t) = e^{-\Gamma t_p^2}$, which is motivated by our observations on Rabi oscillations performed with an ion cooled to the ground state and driven on either the carrier or the motional sideband. We think that this is because these are dominated by shot-to-shot intensity noise. This is consistent with our observation that as we build up data (we typically take many rounds in which only 50 experimental sequences are run per point) the averaged oscillations get smaller as the acquisition time increases, indicating that slow drifts in the Rabi frequency are dominating. Intensity fluctuations would also produce a scaling of the decay parameter with the Rabi frequency, indicating a better choice of dependence for this model of noise is $\gamma(t) = e^{-\Gamma (n + 1) t_p^2}$. That this form does not work so well for cat states probably indicates that decoherence of the cat during the probe pulse is playing a significant role, which we would expect since the expected decoherence times for our heating rate are of similar order to the probe times in the experiment. We performed Monte-Carlo wavefunction simulations with our measured ground state heating rate and motional decoherence which are consistent with this interpretation. 

For all fits used in the Wigner function tomography the decay parameters were extracted using fits performed in the ground state basis (i.e. $\beta = 0$) which includes both the first and second revival. We then fix this value in fits to other values of the displacement, which allows us to use shorter probe times for the displaced bases.

\subsection*{Calibration of displaced and squeezed measurement settings.}
The critical element of the two measurement techniques which we introduce is that they are sensitive to the phase of the quantum state, unlike the energy eigenstate measurement. This requires calibration of the relative phase between the multiple laser fields which are used to generate the Hamiltonians $\hat{H}_d(\beta)$, $\hat{H}_s$ and $\hat{H}_r$. In our experiments, the reference for the phase is the cat state itself, therefore we work with a fixed cat size, and use the predicted behaviour to calibrate the different analysis pulse components.

For $\hat{H}_d(\beta)$, we fix the strength of the carrier term, which sets $|\beta|$, and then scan the phase. The magnitude $|\beta|$ is independently measured by preparing a ground state ion and performing a fit using equation \ref{eq:spinpops} to the Rabi oscillations as a function of time. In this basis the Rabi oscillations are equivalent to applying the blue sideband to an ion prepared in a coherent state and $\ket{\downarrow}$.

For the squeezed basis Hamiltonian $\hat{H}_s$, we scan the relative phase of the red and blue sidebands for a cat of fixed size, and look for a reduction in the Rabi frequency which indicates the basis for which the anti-squeezing is aligned with the cat. The size of the squeezing is calibrated by two methods. First we apply $\hat{H}_s$ to a ground-state cooled ion, resulting in Rabi oscillations which follow \ref{eq:spinpops} with the $p(\ket{\phi_n})$ following the distribution expected for a squeezed state with size given by the Hamiltonian used.

To calibrate the carrier term of $\hat{H}_s(\beta) = \hat{H}_s + \hat{H}_d(\beta)$, we apply this Hamiltonian to a ground-state cooled ion, and observe the change in the dynamics as a function of the carrier phase and amplitude. This references the phase of $\beta$ to the squeezing axis, which is pre-calibrated to the cat phase.

\end{document}